\documentclass[conference, a4paper]{IEEEtran}
\usepackage[left=13mm,right=13mm,top=19mm,bottom=43mm]{geometry}
\usepackage{multirow}
\usepackage{graphicx}
\usepackage{xcolor}
\usepackage{soul}
\usepackage{amssymb}
\usepackage{amsmath}
\usepackage{subcaption}
\usepackage{booktabs}
\usepackage{balance}
\usepackage{microtype}
\usepackage{dblfloatfix}    
\usepackage[utf8]{inputenc}
\usepackage[hidelinks]{hyperref}
\usepackage{csquotes}
\usepackage[american]{babel}
\usepackage{amsmath,amssymb,amsfonts}
\usepackage{mathtools}
\usepackage{graphicx}
\usepackage{xpatch}
\usepackage{braket}
\usepackage{sansmath}
\usepackage[defaultmathsizes, italic]{mathastext}

\usepackage{ifdraft}
\usepackage{etoolbox}

\usepackage[normalem]{ulem}
\usepackage{tabularx}
\usepackage{booktabs}
\usepackage{multirow}
\usepackage{makecell}
\usepackage{textcomp}
\def\nobreakbefore{%
  \relax\ifvmode\else
    \ifhmode
      \ifdim\lastskip > 0pt\relax
        \unskip\nobreakspace
      \fi
    \fi
  \fi
}
\let\oldcite\cite
\renewcommand\cite{\nobreakbefore\oldcite}
\usepackage{enumitem}

\usepackage[bibstyle=ieee, citestyle=numeric-comp, maxbibnames=5, minbibnames=5, maxcitenames=1, mincitenames=1, doi=false, isbn=false, url=false, backend=biber]{biblatex}
\DeclareFieldFormat{journaltitle}{#1\isdot}
\DeclareFieldFormat[book]{title}{{#1\isdot}}

\AtEveryBibitem{
 \clearname{editor}
 \clearfield{month}
  \clearfield{extra}
}
\DeclareSourcemap{
  \maps[datatype=bibtex]{
    \map{    
      \step[fieldset=note, null]
    }
  }
}
\DefineBibliographyStrings{english}{
    andothers = {et al\adddot}
}
\defbibheading{secbib}[\bibname]{
  \section*{#1}%
  \markboth{#1}{#1}}

\newcommand{\updownarrows}{\mathbin\uparrow\hspace{-.3em}\downarrow}
\newcommand{\downuparrows}{\mathbin\downarrow\hspace{-.3em}\uparrow}
\renewcommand{\upuparrows}{\mathbin\uparrow\hspace{-.3em}\uparrow}
\renewcommand{\downdownarrows}{\mathbin\downarrow\hspace{-.3em}\downarrow}

\begin{document}

\title{
SAT-Based Quantum Circuit Adaptation
}

 \author{
Sebastian Brandhofer,$^{1}$ Jinwoong Kim,$^{2}$ Siyuan Niu,$^{3}$ Nicholas T. Bronn$^{4}$
\\\\
\begin{minipage}[c]{\textwidth}
\centering
\small $^1$ Institute of Computer Architecture and Computer Engineering and Center for Integrated Quantum Science and Technology, University of~Stuttgart, Stuttgart, Germany, e-mail: sebastian.brandhofer@iti.uni-stuttgart.de\\
\small $^2$ Applied Physics, Delft University of Technology, Delft, The Netherlands, e-mail: kjw.kim@gmail.com\\
\small $^3$ LIRMM, University of Montpellier, Montpellier, France, e-mail: siyuan.niu@lirmm.fr\\
\small $^4$ IBM Quantum, IBM TJ Watson Research Center, Yorktown Heights, NY, USA, e-mail: ntbronn@us.ibm.com
\end{minipage}

\vspace{-2ex}
}

\maketitle

\begin{abstract}
As the nascent field of quantum computing develops, an increasing number of quantum hardware modalities, such as superconducting electronic circuits, semiconducting spins, trapped ions, and neutral atoms, have become available for performing quantum computations.
These quantum hardware modalities exhibit varying characteristics and implement different universal quantum gate sets that may e.g. contain several distinct two-qubit quantum gates.
Adapting a quantum circuit from a, possibly hardware-agnostic, universal quantum gate set to the quantum gate set of a target hardware modality has a crucial impact on the fidelity and duration of the intended quantum computation.
However, current quantum circuit adaptation techniques only apply a specific decomposition or allow only for local improvements to the target quantum circuit potentially resulting in a quantum computation with less fidelity or more qubit idle time than necessary.
These issues are further aggravated by the multiple options of hardware-native quantum gates rendering multiple universal quantum gates sets accessible to a hardware modality.
In this work, we developed a satisfiability modulo theories model that determines an optimized quantum circuit adaptation given a set of allowed substitutions and decompositions, a target hardware modality and the quantum circuit to be adapted.
We further discuss the physics of the semiconducting spins hardware modality, show possible implementations of distinct two-qubit quantum gates, and evaluate the developed model on the semiconducting spins hardware modality.
Using the developed quantum circuit adaptation method on a noisy simulator, we show the Hellinger fidelity could be improved by up to 40\% and the qubit idle time could be decreased by up to 87\% compared to alternative quantum circuit adaptation techniques.

\end{abstract}

\maketitle

\section{Introduction}

Currently, numerous noisy quantum hardware modalities for quantum computing satisfy the requirements for universal quantum computing. However, no as-realized hardware modality has achieved fault-tolerance. Therefore, while universal, current quantum modalities are  limited in circuit depth by gate fidelity and ultimately by the qubit coherence times. Therefore circuit adaptation strategies that reduce gate count, transform operations into equivalent ones with less incurred error, and reduce qubit idle time, are highly relevant for quantum advantage on near-term noisy quantum hardware.

An arbitrary quantum computation may be decomposed into a set of single- and two-qubit gates~\cite{14, 17}, however hardware modalities typically suffer higher infidelity from the two-qubit gates than single-qubit. Hence quantum circuit adaptation for hardware modalities that admit multiple two-qubit gates can be used to improve overall result quality. Here we choose the modality of spin qubits in semiconductor dots as an example that admits three two-qubit gates depending on parameter regime and control to study a satisfiability modulo theories (SMT) model for quantum circuit adaptation.
\par

In this work, we begin with a discussion of the physics of spin qubits in semiconducting quantum dots in section~\ref{4}. Depending on the control applied, each of three two-qubit gates swap, CPHASE, and CROT can be realized.
The latter two of which are universal together with single-qubit rotations while the swap operation is useful for limited-connectivity modalities such as spins.

This is followed by a discussion of the costs associated with each operation, such as infidelity and duration, that will inform our SMT model. After an overview of circuit adaptation techniques in section~\ref{6}, we provide a rigorous description of our SMT model and an example for illustration in section~\ref{9}. 

\section{Two-qubit Gates in Semiconductor Spin Qubits}\label{4}
We first demonstrate how to realize two-qubit gates from an effective Hamiltonian based on the physics of spin qubits. A summary of the differing fidelities and durations of such operations follows, preparing the case for our satisfiability modulo theories model.

In general, quantum dots are formed by isolating electrons used as qubits from other electrons in the reservoir. In a two-dimensional electron gas, separation is achieved by applying voltages to barrier and plunger gates, which deplete charge carriers in the semiconducting layer. Manipulation of qubits is also achieved via the barrier and plunger gates~\cite{10}. \par

\subsection{Dynamics of Two-qubit Gates} \label{sec:two-qubit-ctrl}

\begin{figure}[tb]
    \centering
    \includegraphics[width=0.42\textwidth]{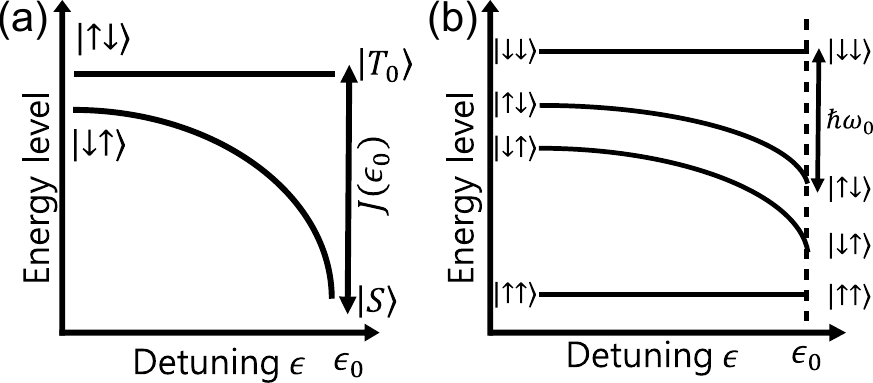}
    \caption{\textbf{Eigenenergies in different regimes}.(a) $J(\epsilon)>>\Delta E_z$. Used for swap gate protocol. (b) $\Delta E_z >> J(\epsilon)$. Used for CROT/CPHASE gate protocols.   \vspace{-3ex}
}
    \label{3}

\end{figure}

Eigenstates and eigenvalues of an effective Hamiltonian dictate the dynamics of a two-qubit system. These are depicted in Fig. \ref{3}a and Fig. \ref{3}b for different parameter regimes. Here, the effective Hamiltonian leads to three widely-used two-qubit gates: the swap gate, controlled phase gate (CPHASE), and controlled rotation gate (CROT). Note that $\pi$-angle rotations for CPHASE and CROT result in CZ and CNOT gates, respectively.\par

Following the proposal in \cite{12}, Swap gates can be realized by harnessing natural spin exchange interaction, which is most pronounced when the exchange splitting dominates the magnetic field gradient, i.e. $J(\epsilon) \gg \Delta E_z$. Energy eigenvalues of the effective Hamiltonian are shown in Fig. \ref{3}a. As detuning increases, the eigenstates switch from $|\updownarrows\rangle$ and $|\downuparrows\rangle$ to $|T_0\rangle = |\updownarrows\rangle + |\downuparrows\rangle$ and $|S\rangle = |\updownarrows\rangle - |\downuparrows\rangle$ (up to normalization), respectively. At detuning $\epsilon_0$, the exchange splitting $J(\epsilon_0)$ between eigenstates are large enough to induce two-qubit operation within coherence time. Swap gate consists of three control signals: (1) increase detuning $\epsilon_0$ diabatically preserves the initial state, for example $|\downuparrows\rangle = |T_0\rangle - |S\rangle$ in the new eigenbasis, (2) precess the initial state in a projected two-qubit Bloch sphere defined by $|S\rangle$ and $|T_0\rangle$ for time $\tau_{op} =\pi/J(\epsilon_0)$ corresponding  to a $\pi$-rotation in the $z$-axis, and (3) decrease detuning diabatically to the initial configuration. \par

Common choices for two-qubit entangling gates for spin qubits are CPHASE, which changes the phase of the target qubit depending on the state of the control qubit, and CROT, which rotates the target qubit based on the state of the control qubit. Both quantum gates can be realized in quantum dots when the magnetic field gradient is much larger than exchange splitting, i.e. $\Delta E_z \gg J(\epsilon)$. In this case, energy eigenvalues of the effective Hamiltonian either shift or stay the same depending on eigenstate. When eigenstates are anti-parallel, eigenenergies shift in energy with increasing $\epsilon$, while for parallel eigenstates, they remain the same as shown in Fig. \ref{3}b. In all cases, the eigenstates remain in their initial states. Implementation of CPHASE utilizes the difference in eigenenergies of anti-parallel states before and after applying detuning $\epsilon$. When an adiabatic pulse is applied to the system (keeping the system in an eigenstate of the Hamiltonian), anti-parallel states accumulate phases relative to parallel states, which is equivalent to CPHASE up to single-qubit gates.\par

Implementation of CROT takes advantage of phase shifts of anti-parallel states in a different way. As shown in Fig. \ref{3}b, when the system is adiabatically pulsed with the detuning $\epsilon_0$, the shifts in eigenenergies of anti-parallel states cause transition frequencies between $|\downdownarrows\rangle$-$|\updownarrows\rangle$ and $|\downuparrows\rangle$-$|\upuparrows\rangle$ to deviate from each other. As we have different resonant frequencies for each transition~\cite{6}, we can separately drive the desired transition. This equips us with CROT in our native quantum gate set. Note that the use of adiabaticity in CPHASE and CROT may depend on the used control schemes and the underlying material.
\subsection{Different Implementations of Two-Qubit Gates\label{sec:spin_gates}}

While the introduced two-qubit gates are in principle possible, there are numerous difficulties in exploiting the potential of having different two-qubit gates in one platform. First, preferences of some two-qubit gates for device characteristics are in contrast with those of other two-qubit gates. Second, gates require adiabatic pulses to suppress unwanted rotations caused by residual interactions. Adiabatic control allows the eigenstates to remain in their original states by the adiabatic theorem, thereby resulting in high gate fidelities. However, adiabatic  control typically increases the gate operation time. Hence, the spin-qubit platform can have higher gate fidelity two-qubit gate realizations with a longer gate time and two-qubit gate realizations with a lower gate fidelity but also shorter gate time. As we show, the characteristics of such realizations could be exploited to increase the overall circuit fidelity. Third, residual interactions occasionally persist despite the adiabatic pulse and degrade gate fidelities. For example, if one performs a swap in the presence of a specific Zeeman energy difference $\Delta E_z$, an unwanted rotation caused by $\Delta E_z$ will deteriorate the performance of the swap operation.

These challenges have been tackled by numerous papers \cite{15, 8, 13} to either increase the fidelity or decrease the gate time. Notably, \cite{15} implemented all two-qubit gates discussed above with high fidelities and short gate times within a single spin-qubit platform. The authors used a geometric gate to decrease gate time and increase gate fidelity. Also, they utilized composite pulses to further suppress unwanted rotations. As a result, they achieved gate times $D_{0}$ in table \ref{7}. However, a scaled-up spin-qubit platform may require different materials or driving mechanisms. We therefore also investigate gate time $D_{1}$ with the fidelities depicted in table~\ref{7}, where CZ$_\text{db}$ is a diabatic CZ gate, SWAP$_\text{d}$ is a diabatic swap gate and SWAP$_\text{c}$ is a swap gate realized by composite pulses.
\begin{table}[]
\setlength{\tabcolsep}{5.4pt}
\caption{\label{7}Investigated gate durations and fidelities}
\begin{tabular}{@{}lccccccc@{}}
                  & SU(2) & CZ  & CZ$_\text{db}$  & CROT   & SWAP$_\text{d}$ & SWAP$_\text{c}$   \\ \midrule
Fidelity          & 0.999& 0.999 & 0.99 & 0.994 & 0.99 & 0.999  \\
Duration $D_{0}$ {[}ns{]} & 30 & 152 & 67   & 660  & 19    & 89     \\
Duration $D_{1}$ {[}ns{]} & 30 & 151 & 7   & 660  & 9    & 13     \\
\end{tabular}
\vspace{-3ex}
\end{table}

\section{Circuit Adaptation Techniques}\label{6}

Quantum computing can be realized on different quantum modalities with distinct hardware limitations. Each quantum technology has its specific basis gate sets that consist of entangling gates and single-qubit gates. For example, the basis gate sets of IBM quantum hardware include a single two-qubit gate (CNOT), while for spin qubit devices, the target of this paper, the entangling gates are typically CPHASE or CROT. However, a quantum circuit is usually generated for an abstract gate set or may have been generated for a different hardware modality than the actual target hardware modality. Therefore, it needs to be adapted to the given quantum hardware containing only the basis gate sets. Here, we introduce several commonly used circuit translation techniques.

\textbf{Direct Basis Translation.}
This technique translates the quantum gates from the source basis defined by the input circuit to the target basis according to a pre-defined {\it equivalence library}. The equivalence library includes various ways of decomposing a gate to its equivalent implementations. For example, a CNOT gate can be decomposed equivalently to a set of single-qubit gates in conjunction with one of the three different two-qubit basis gates: CZ, iSWAP, or $R_{\text{zx}}$. 
If these gates occur in a target quantum circuit, they can each be replaced by the CNOT basis gate and single-qubit gates as defined in the equivalence library and its gate substitution rules. 
    
\textbf{Template Optimization.}
This is a circuit optimization technique, typically used for reducing the error or duration of quantum operations~\cite{0zit}, that consists of three individual steps. First, the template input to the technique must be constructed.
A template is generally defined to be a quantum circuit that evaluates the identity operation.
The template consists of two parts that are functionally inverses but typically have different basis gates, some examples of which are shown in Figure.~\ref{8}(a)-(d).
In the second step, template matching is performed, which aims at finding all the maximal matches of the input templates in the target circuit~\cite{7,1,2}.
In a final step, template substitution is performed. During this step, the matched part of the original subcircuit is replaced by the inverse of the unmatched part if the unmatched part of the template has a lower cost. The cost can be evaluated with various metrics, such as gate implementation cost, error rate, or gate duration. As opposed to direct basis substitution, where non-basis gates are simplified by targeting basis gates through the equivalence library, template optimization offers the flexibility of converting between different basis gates and optimizing certain circuit patterns more effectively.

\textbf{Unitary Decomposition.}
This is the process of translating a given unitary matrix to a sequence of single and two-qubit gates. This is also known as circuit synthesis. It can be particularly useful for applications composed of arbitrary unitary gates, such as quantum volume circuit~\cite{4}, to convert the unitary matrices to hardware basis gate sets and generate a synthesized circuit. Several methods have been proposed to reduce the number of gates in the synthesized circuit, such as cosine-sine matrix decomposition (CSD)~\cite{5}, quantum Shannon decomposition (QSD)~\cite{3}, and KAK decomposition~\cite{11}.

\textbf{Suitability for Quantum Circuit Adaptation.}\label{sec:related_work}
Various circuit adaptation techniques introduced in this section are defined as transpilation passes in a quantum compiler and each of them works well independently. However, if each technique is applied separately during the circuit transpilation process, the performance of the quantum circuit adaptation is limited.

For direct basis translation and the unitary decomposition method, the adaptation is only performed with one two-qubit basis gate which lacks flexibility when a combination of both would improve the quantum circuit even further. While it is possible for template optimization to adapt a quantum circuit to various two-qubit basis gates, only a local solution can be determined for one template at a time~\cite{16}. The same result quality as possible with a global optimization relying on evaluating multiple templates at the same time can not be reached. A clear method for combining these approaches in an optimized way remains elusive, and is the subject of our investigation. Our proposed method incorporates the above approaches in a circuit adaptation technique such that an adapted quantum circuit with high fidelity is obtained. The variations obtained are specifically evaluated for translating a quantum circuit to a spin-qubit device with multiple two-qubit basis gates.

\section{SAT-Based Quantum Circuit Adaptation}\label{9}
\begin{figure}[htbp]
  \centering
  \includegraphics[width=1.0\linewidth]{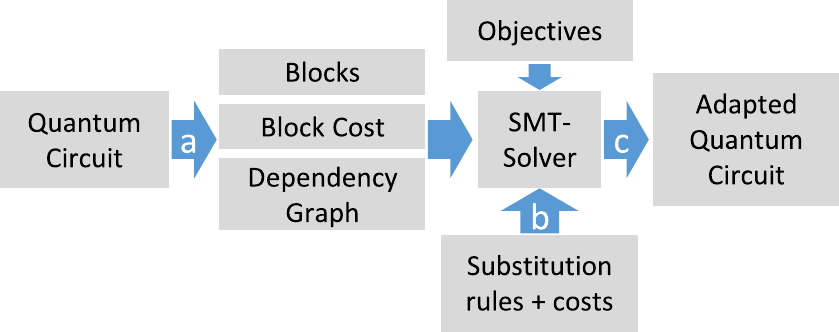}
  \caption{Workflow of the developed quantum circuit adaptation method for an arbitrary input quantum circuit with preprocessing steps (a), substitution rule evaluations (b) as well as SMT model construction and SMT solving (c).\label{5}}
\end{figure}

The steps of the proposed method for adapting a quantum circuit from one quantum hardware modality or an abstract gate set to a target quantum hardware modality are shown in Fig.~\ref{5}.
First, the quantum circuit is preprocessed to yield a set of blocks along with their dependencies and their cost in terms of fidelity and duration.
Then, every specified substitution rule is evaluated on the quantum circuit.
The preprocessed quantum circuit, the specified substitution rules and the defined objective function are used to construct an SMT model in a third step.
The SMT model is then input to a SMT solver~\cite{18}~that computes an assignment to the model variables such that the objective function is optimized. The assignment is then used to derive an adapted quantum circuit using a selection of specified substitution rules.

The following sections describe the applied preprocessing steps as well as the evaluation of specified adaptations, then show the construction of the SMT model and give an explicit example for adapting a quantum circuit designed for IBM quantum computers~\cite{19}~to the spin qubit hardware modality specified in~\cite{15,23}.
\subsection{Preprocessing}\label{sec:preprocessing}
Preprocessing consists of three steps that are applied successively.
First, the input quantum circuit is partitioned into two-qubit \emph{blocks} that contain gates interacting on the same qubit pair.
The order of the blocks is given by a block \emph{dependency graph} that contains each block $b$ as a vertex and an edge $a=(b', b)$ if block $b'$ must be computed before block $b$.

In a second step, each basis gate of the source quantum hardware modality is substituted by basis gates of the target hardware modality.
The basis gate substitution can be performed using an equivalence library that can be generated manually~\cite{21}~or automatically~\cite{20}.

Finally, the cost of each block after basis gate substitution is evaluated in terms of block duration and block fidelity.
The block duration is the length of the critical path in the block, i.e. the time the target quantum computer needs to execute the block.
The block fidelity is defined as the product of each gate fidelity in the block.
Basis gate translation provides a naive adaptation that is used as a common reference cost in subsequent steps.

\subsection{Evaluation of Substitution Rules}\label{sec:subrules}

Each specified substitution rule is evaluated on the input quantum circuit and then 
used to define an SMT model in a subsequent step.
During the evaluation of a substitution rule, the set of substituted gates $p_s$, the set of substitution gates $g_s$, the affected blocks $b_s$ and the cost of the substitution $w_s$ are determined for each substitution $s$ applicable to the quantum circuit.

A substitution rule can be a gate equivalency, quantum circuit equivalency or a decomposition method that decomposes a block to the basis gates of the target hardware modality.
The substitution rules can be defined manually by a domain expert as a set of quantum circuit or gate equivalencies~\cite{16}, derived automatically~\cite{20}~for the basis gates of the target hardware modality, or be part of a general decomposition method such as the KAK decomposition~\cite{11}.

Quantum circuit or gate equivalency substitution rules can be evaluated in polynomial runtime~\cite{16}. 
Evaluating substitution rules based on decomposition requires one to first compute the unitary matrix of each block and then evaluate the cost of a decomposition.
Determining the unitary matrix of $n$-qubit block requires a runtime exponential in the number of qubits $n$.
However, for small $n$, in our case $n=2$, the runtime overhead is not significant.

\subsection{SMT Model for Quantum Circuit Adaptation}

In this section we describe how the data from the preprocessing steps and the substitution rule evaluation are used to generate an SMT model that yields a quantum circuit adaptation from a source hardware modality to a target hardware modality.
The developed SMT model consists of Boolean variables, constraints and the definition of linear objective functions.
An SMT solver computes an assignment to the variables of the developed SMT model that is satisfying all constraints and that is optimal with respect to the defined model assumptions.
In this work the Z3 solver software was used as an SMT solver~\cite{18}.

\subsubsection{SMT Model Variables}
The developed SMT model for a quantum circuit with $S$ substitutions, $B$ blocks and a dependency graph $G = (V, A)$ with vertices $V$ and edges $A$ consists of variables:
\begin{itemize}
    \item $C = \{c_{1}, ..., {c_{|S|}}\}$: the set of chosen substitutions for the quantum circuit adaptation, i.e. the resulting quantum circuit adaptations only contain a substitution $s$ if $c_{s}$ evaluates to true.
    \item $E = \{e_{1}, ..., {e_{|B|}}\}$: the set of block starting times, i.e. the time at which the computation of a block is started on the target hardware modality.
    \item $D = \{d_{1}, ..., {d_{|B|}}\}$: the set of block duration times
    
    \item $F = \{f_{1}, ..., {f_{|B|}}\}$: the set of block fidelity
    
\end{itemize}
\subsubsection{SMT Model Constraints}
The assignment to sets $C, E$ and $D$ must be constrained to yield a valid and optimized quantum circuit adaptation.
First, a substitution may only be chosen in a quantum circuit adaptation, if it does not substitute the same gates as another chosen substitution:
\begin{equation}
    \neg c_{s} \vee \neg c_{s'} \qquad \forall s, s' \in S: p_{s} \cap p_{s'} \neq \emptyset,
\end{equation}
where $p_{s}$ and $p_{s'}$ are the sets of quantum gates that will be substituted by substitutions $s$ and $s'$, respectively.
The symbol $\neg$ refers to the logic negation while the symbol $\wedge$ ($\vee$) corresponds to the logic conjunction (disjunction).
In addition, the computation of a block in a quantum circuit must obey the dependency defined in graph $G$.
Thus, the computation of a block $b$ on a target quantum computer may only start if the computation of any preceding block $b'$ has been concluded:
\begin{equation}
    e_{b} \geq e_{b'} + d_{b'} \qquad \forall b, b' \in B: a_{b', b} \in A,
\end{equation}
where $a_{b', b}$ is an edge in the block dependency graph $G$, $e_{b'}$ is the time step at which the computation of block $b'$ and $d_{b'}$ is the duration of computing block $b'$.
Finally, the block duration time and block fidelity must be set depending on the chosen substitutions in the quantum circuit adaptation.
The block duration time $d_{b}$ of a block $b$ is set by:
\begin{equation}
    d_{b} := D(b) + \sum_{s\in S'} \mathbb{D}(s) \wedge c_{s},    
\end{equation}
where ${D}(\cdot)$ returns the duration of a block or quantum gate, and $\mathbb{D}(\cdot)$ gives the reduction in duration incurred by a substitution.
The duration reduction is defined by
\begin{equation}
    \mathbb{D}(s) = \sum_{g\in g_{s}} D(g) - \sum_{p\in p_{s}} D(p),
\end{equation}
where $g_s$ is the set of substitution quantum gates and $p_s$ is the set of substituted quantum gates of substitution $s$.
Likewise, the fidelity $f_{b}$ of a block $b$ is determined by:
\begin{equation}\label{13}
    f_{b} := \log({F}(b)) + \sum_{s\in S} \mathbb{F}(s) \wedge c_{s}    
\end{equation}
where $F(\cdot)$ returns the fidelity of a quantum gate or of a block given by the reference adaptation determined during preprocessing steps, and $\mathbb{F}(\cdot)$ gives the improvement in fidelity incurred by a substitution.
The improvement in fidelity is defined by:
\begin{equation}
    \mathbb{F}(s) = \sum_{g\in g_{s}} \log(F(g))-\sum_{p\in p_{s}} \log(F(p)),
\end{equation}
where $g_s$ are the substitution quantum gates and $p_s$ are the substituted quantum gates of substitution $s$.

Note that the developed model does not contain functions ${D}(\cdot)$ and $\log({F}(\cdot))$. Instead, the function value of every substitution $s$, quantum gate $g$ and block $b$ in the reference adaptation is computed before the SMT model is constructed.
Furthermore, the developed model only registers one duration and start time for a \emph{two-qubit} block.
This introduces single-qubit gate ambiguities when minimizing the qubit idle time or quantum circuit duration if the duration (a single-qubit gates) on one qubit is different to the other in a template or block.

\subsubsection{Objective Functions}
Lastly, we describe the objective functions investigated in this work. 
The objective function provided to the SMT solver is crucial for improving the quantum circuit adaptation, i.e. improving the probability of computing the correct result on a noisy, near-term hardware modality.
We investigated objective functions that improve the fidelity of the adapted quantum circuit, qubit idle time of the adapted quantum circuit and a combination of both.
The qubit idle time has been observed to be a source of error~\cite{22}~that should be minimized in a quantum circuit.
We assume the state of a qubit to decay during idle time, i.e. the state of a qubit is unaffected by the idle time with probability:
\begin{equation}
    e^{-d/T},
\end{equation}
where $d$ is the duration during which a qubit is idle and $T$ is the coherence time of the target hardware modality.
The fidelity objective of the adapted quantum circuit is defined by:
\begin{equation}\label{1}
    \max \sum_{b} f_b,
\end{equation}
where $f_b$ is defined as in Eq.~\ref{13}. The qubit idle time in the adapted quantum circuit is optimized by:
\begin{equation}\label{0}
    \max \; -\frac{Q\cdot \text{D} - \sum_{b} d_{b}}{T},
\end{equation}
where $Q$ is the number of qubits and D is the total circuit duration.
We also combine these objectives as a product:
\begin{equation} \label{12}
\begin{split}     
      &\max \sum_{b} \log(f_{b}) - \frac{Q\cdot D - \sum_{b} d_{b}}{T}.     
\end{split}
\end{equation}

\subsubsection{Determining the SMT Quantum Circuit Adaptation}

After an SMT solver computed an assignment to the SMT model variables, a substitution $S$
is applied to the target quantum circuit if $c_s$ is set by the SMT solver.
A substitution $s$ is applied to a quantum circuit by substituting quantum gates $p_s$ in the quantum circuit with $g_s$.
A quantum gate in the original quantum circuit is substituted by the basis translation performed in the preprocessing step if the quantum gate is not part of any chosen substitution.

\subsection*{Example: Adapting a Quantum Circuit from IBM Backends to Spin Qubits}
\begin{figure}[htbp]
  \centering
  \includegraphics[width=1.0\linewidth]{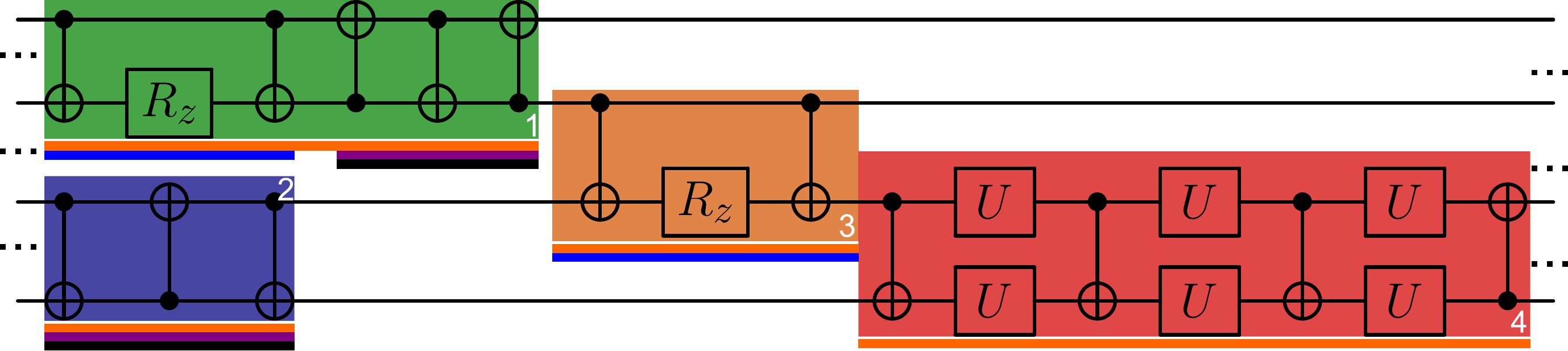}
  \caption{Quantum circuit adaptation for an example quantum circuit given in the IBM backend basis. Continuous lines indicate quantum gates substituted by the same substitution rule. An orange line corresponds to a KAK decomposition, a blue line to conditional rotation gates, and violet and black lines to different swap gate realizations. \label{11}}
\end{figure}

In this section we describe the adaptation of a quantum circuit given in the basis of an IBM quantum computer based on superconducting qubits~\cite{19}~to a quantum circuit suitable for computation on a spin-qubit quantum computer~\cite{15}.
Figure~\ref{11}~shows the quantum circuit and table~\ref{7}~($D_{0}$)~shows the characteristics of the quantum gates supported by the spin-qubit quantum computer used in this example~\cite{15}.
The corresponding spin-qubit quantum computer supports arbitrary single-qubit gates in SU(2), a two-qubit controlled-Z (CZ) gate that is also used for KAK decompositions, two-qubit conditional rotation gates along an arbitrary axis
and two native realizations of swap gates (swap$_\text{d}$ and swap$_\text{c}$).
We do not consider the diabatic CZ gate in this example.
The swap gate realization swap$_\text{d}$ requires less time than the swap gate swap$_\text{c}$ but also has a lower gate fidelity than swap$_\text{c}$.
Depending on the structure of the quantum circuit, the swap gate swap$_\text{d}$ or the swap gate swap$_\text{c}$ may be preferable in a quantum circuit adaptation, e.g. for reducing qubit idle time.
\begin{figure}[htpb]
  \centering
\begin{tabular}[c]{cc}
    \begin{subfigure}[c]{0.5\linewidth}
      \includegraphics[height=0.8cm]{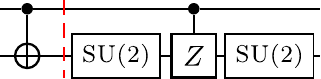}
      \caption{(diabatic) Conditional-Z}
      \label{14}
    \end{subfigure} &
    \hspace{-6ex}
    \begin{subfigure}[c]{0.5\linewidth}    
      \includegraphics[height=0.8cm]{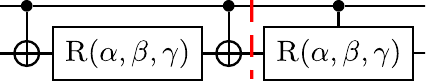}
      \caption{Conditional-Rotation (CR)}
      \label{fig:b}
    \end{subfigure} \\
    
    \begin{subfigure}[c]{0.5\linewidth}
    \vspace{1ex}
      \includegraphics[height=0.8cm]{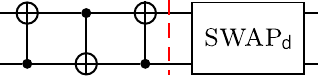}
      \caption{Direct swap gate}
      \label{fig:c}
    \end{subfigure} &
    \begin{subfigure}[c]{0.5\linewidth}
    \vspace{1ex}
      \includegraphics[height=0.8cm]{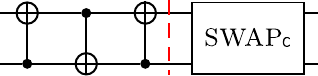}
      \caption{Composite swap gate}
      \label{fig:d}
    \end{subfigure} \\
\end{tabular}
\begin{subfigure}[c]{\linewidth}
    \vspace{1ex}
    \centering
      \includegraphics[height=1cm]{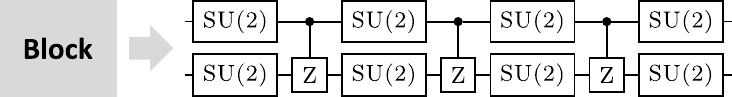}
      \caption{KAK decomposition using CZ and single-qubit gates}
      \label{fig:e}
    \end{subfigure} 
\caption{Substitution rules for adapting quantum circuits generated for IBM backends~\cite{19}~to spin-based systems~\cite{9}\vspace{-2ex}
}
\label{8}
\end{figure}

The results of the quantum circuit adaptation steps are shown in figure~\ref{11}.
First, the target quantum circuit is partitioned into blocks and the basis gate translation (see figure~\ref{14}) is performed to determine a reference cost for each block.
The substitution rules described in figure~\ref{8}~are evaluated on the target quantum circuit in the next step.
This yields ten substitution matches, where the KAK decomposition (orange line) could be applied once to each block, the conditional-rotation (blue line) could be applied in block 1 and block 3 once, and swap$_\text{d}$ (violet line) as well as swap$_\text{c}$ (black line) could each be applied once in block 1 and block 2.
The duration of block 1 is set in our example by:
\begin{multline}
    d_{1} = 965 + (573-965) \wedge c_{0} + (660 - 422) \wedge c_{1} + (19 - 543)\wedge c_{2}\\ + (67 - 543)\wedge c_{3},
\end{multline}
where $965$ns is the reference block duration given by the basis translation, $c_{0}, c_{1}, c_{2}, c_{3}$ corresponds to whether the KAK decomposition $(c_{0})$, the conditional-rotation substitution $(c_{1})$, the direct swap substitution swap$_d$ $(c_{2})$ or the composite swap substitution swap$_c$ $(c_{3})$ is applied.
Characteristics of the other blocks are computed in an analogous way and input to the SMT model construction (see section~\ref{9}).
Depending on the chosen objective function different substitutions may be applied during the quantum circuit adaptation.
In this example we assume that the quantum circuit duration should be minimized.
Using a KAK decomposition, the duration of block 1 would be reduced by 392ns, the conditional-rotation quantum gate would increase the duration by 238ns, swap$_\text{d}$ reduces the duration by 524 and swap$_\text{c}$ reduces the duration by 476ns.
Substitutions $s_0$, $s_2$ and $s_3$ as well as substitutions $s_0$ and $s_1$ are incompatible since they substitute the same set of quantum gates.
Thus, applying KAK substitution $s_0$ reduces the duration of block 1 the most.

The values and equations for the block duration and dependency are entered as an SMT model into an SMT solver whose result informs the quantum circuit adaptation.

\section{Results}\label{sec:results}
In this section, we evaluate the developed SMT model on the introduced semiconducting spin hardware modality.
We investigated the increase in circuit and Hellinger fidelity, and decrease in qubit idle for quantum volume circuits~\cite{4}~and random circuits containing gates from the templates in Fig.~\ref{8}~with up to 4 qubits and a quantum circuit depth of up to 160.
Two gate characteristics $D_{0}, D_{1}$ as given in table~\ref{7}~were evaluated.
The developed SMT model is compared to employing a KAK decomposition using CZ and diabatic CZ gates, template optimization with two objectives targeting the quantum circuit fidelity and qubit idle time, and a direct basis translation that replaces each non-supported two-qubit quantum gate in the quantum circuit with a CZ gate.
The SMT solver was invoked with the fidelity objective $\text{SAT } F$ given in Eq.~\ref{1}, the idle time objective $\text{SAT } R$ given in Eq.~\ref{0}, and the combined objective $\text{SAT } P$ as given in Eq.~\ref{12}.
The quantum circuit determined by direct basis translation is chosen as a baseline for comparison in the following results.
Before employing the quantum circuit adaptation technique, Qiskit~\cite{21}~was used to transpile the target quantum circuit into one compliant with the hardware topology. 

\subsection{Circuit Fidelity Increase and Qubit Idle Time Decrease}

\begin{figure}[t!]
  \centering
  \includegraphics[width=0.75\linewidth]{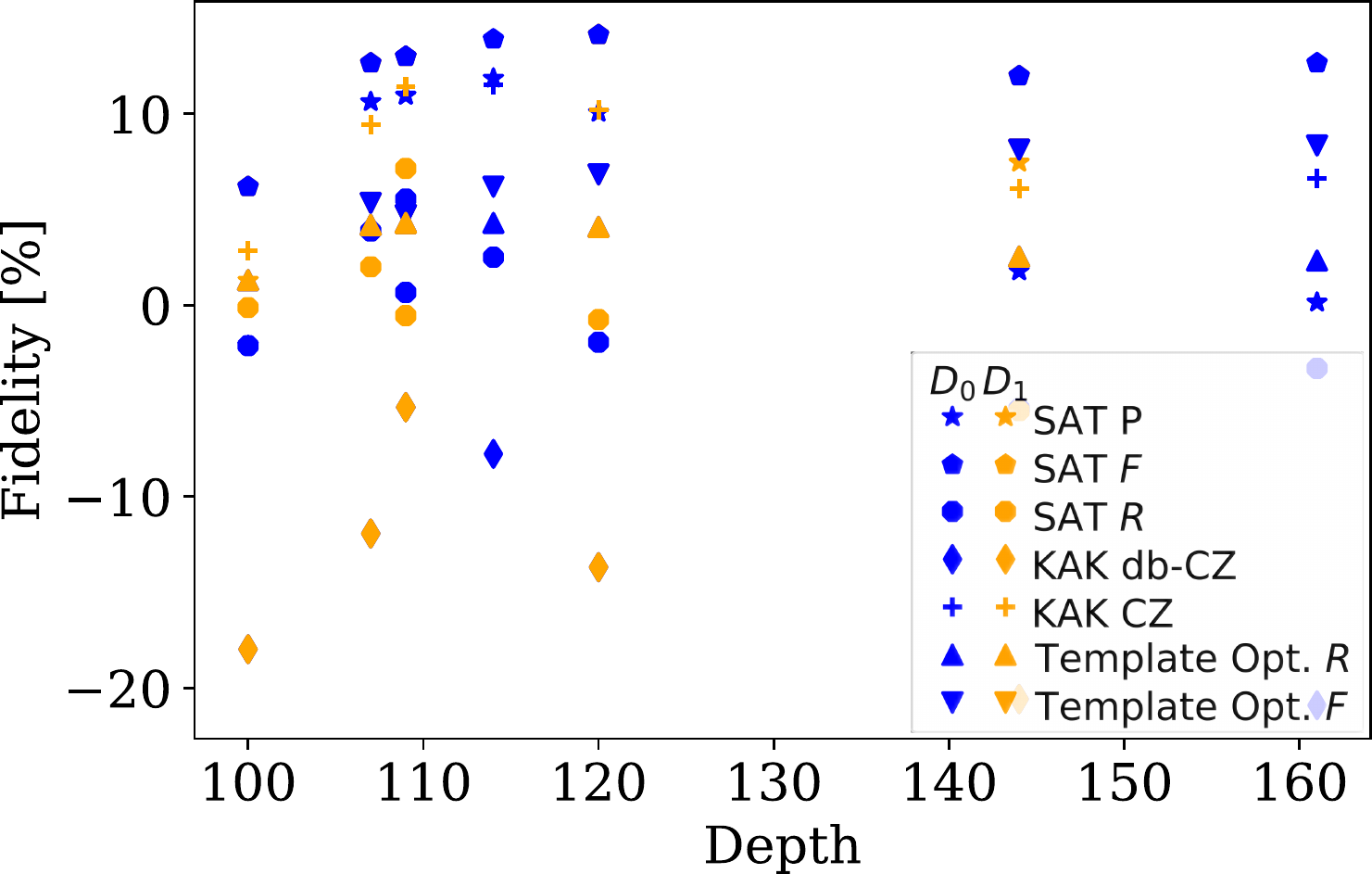}
  \caption{Change in quantum circuit fidelity as given by the product of gate fidelities.\label{2}    \vspace{-2ex}
}
\end{figure}

\begin{figure}[t!]
  \centering
  \includegraphics[width=0.75\linewidth]{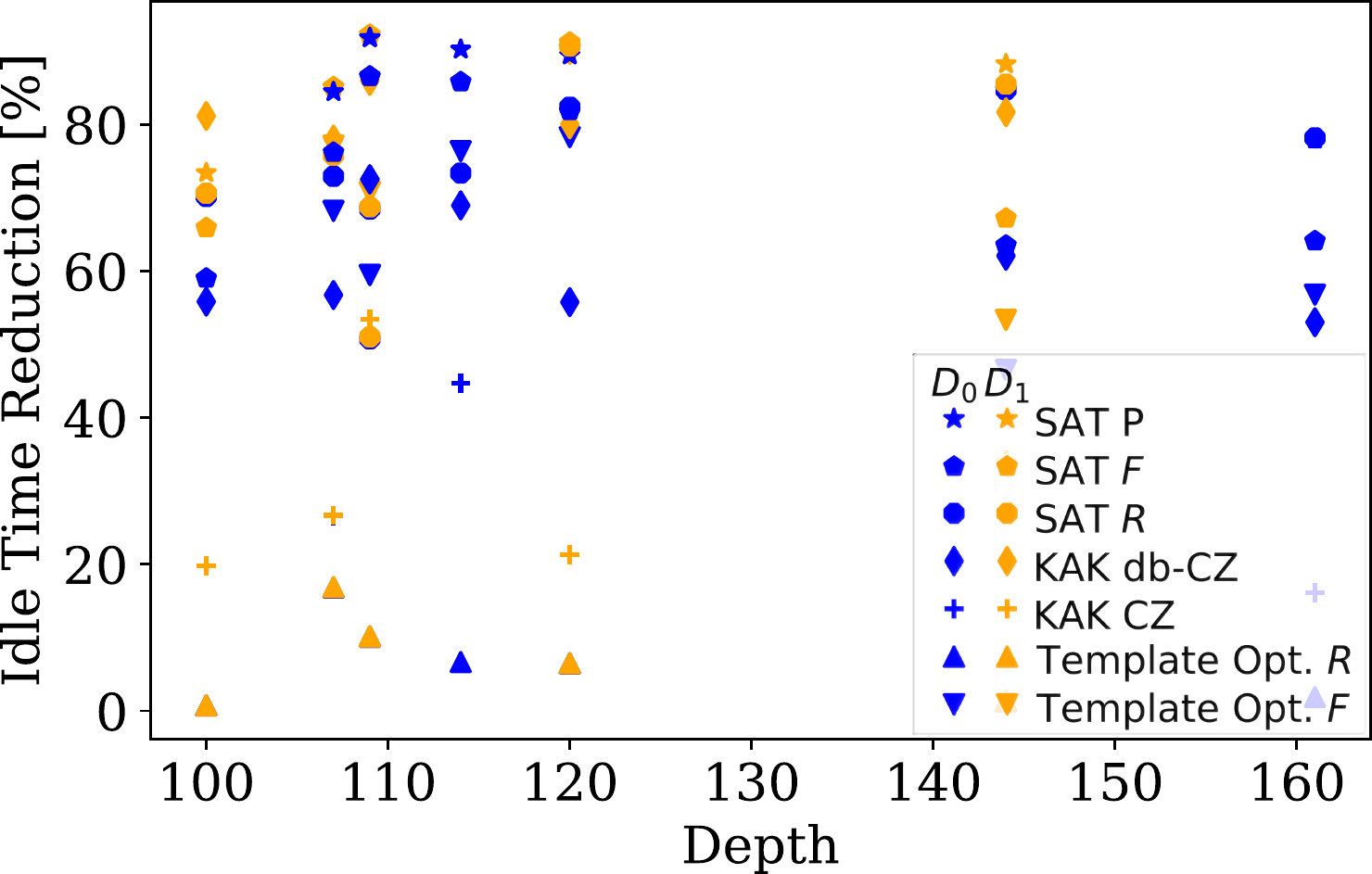}
  \caption{Decrease in qubit idle time yielded by the analyzed quantum circuit adaptation techniques.\label{10}}
\end{figure}

In this section, we evaluated the impact of quantum circuit adaptation on the decrease in qubit idle time, and the change in quantum circuit fidelity as given by the product of individual gate fidelities.
The fidelity and idle time of the quantum circuit as determined by direct basis translation is chosen as a baseline for comparison in the following results.
As depicted in Fig.~\ref{2}~the SMT approach yields the largest improvement in quantum circuit fidelity of up to 15\% over all quantum circuits.
Performing quantum circuit adaptation by only using KAK decompositions based on (diabatic) CZ gates decreases the overall quantum circuit fidelity since the KAK decomposition may introduce additional single-qubit gates compared to template optimization.
In addition, the diabatic CZ gate has a lower gate fidelity as the baseline basis translation using CZ gates (see table~\ref{7}).
In figure~\ref{10}~the decrease in qubit idle time of the respective quantum circuits is depicted for the studied quantum circuit adaptation techniques.
The SMT based approaches yield the highest decrease in qubit idle time for all but the smallest quantum circuit.

\subsection{Hellinger Fidelity and Qubit Idle Time}
Here, we investigate the impact of the developed approach on the qubit idle time and the Hellinger fidelity obtained by performing quantum circuit simulation subject to errors incurred by a depolarization channel that corresponds to the individual gate fidelities and thermal relaxation that corresponds to the qubit idle time~\cite{21}.
In accordance to~\cite{15}, we assumed $T_{2}=2900$ ns and a $T_{1}$ time that is three orders of magnitudes larger for thermal relaxation errors. 
\begin{figure}[htbp]
  \centering
  \includegraphics[width=0.75\linewidth]{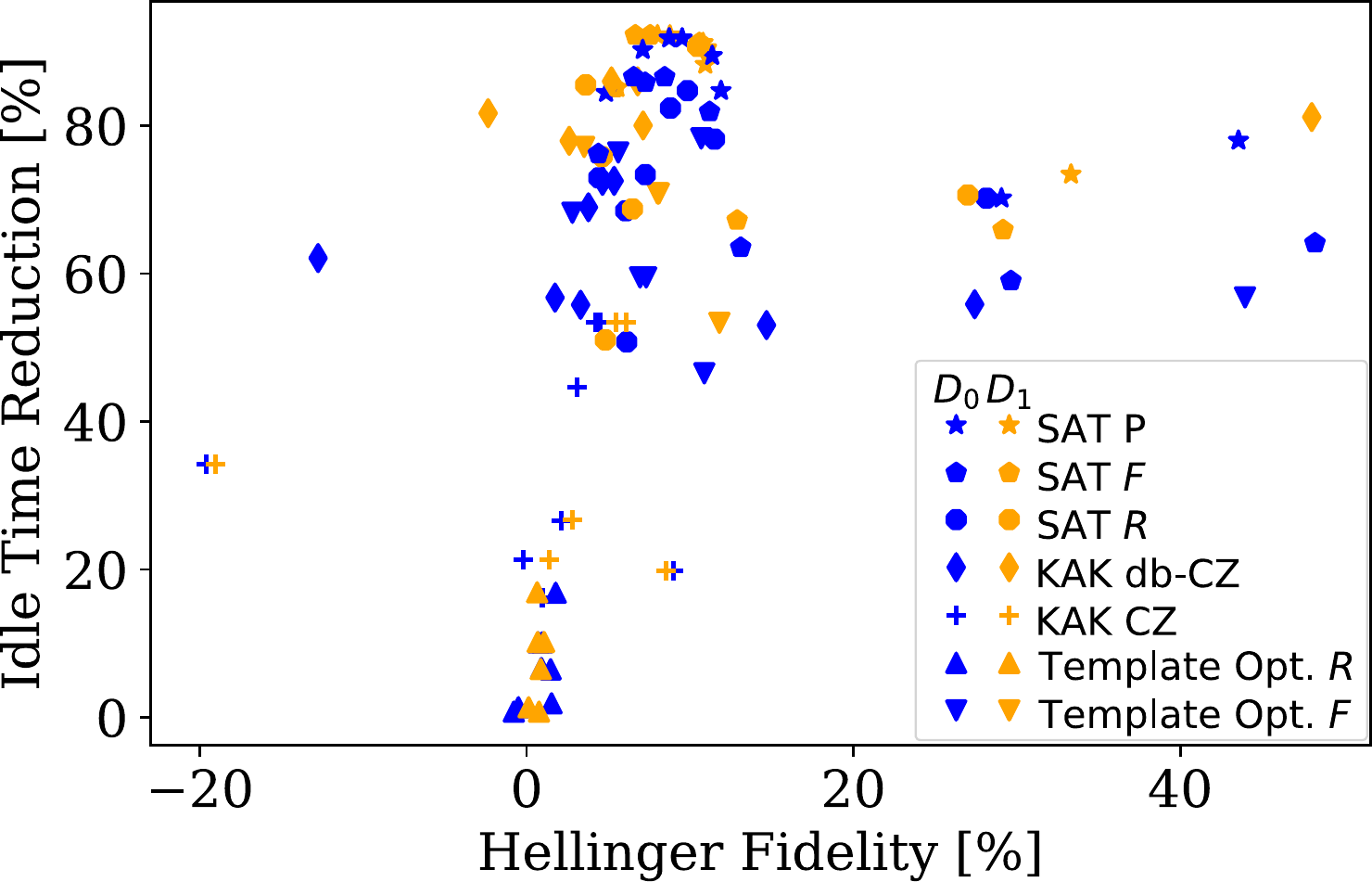}
  \caption{Change in Hellinger fidelity and qubit idle time for the studied quantum circuit adaptation techniques.\label{fig:rt_fid}}
\end{figure}
Figure~\ref{2}~shows the decrease in qubit idle time on the y-axis and the change in Hellinger fidelity on the x-axis.
The developed SMT approaches yield adapted quantum circuits with the highest decrease in qubit idle time and the largest increase in Hellinger fidelity.
The evaluated quantum circuit adaptation techniques based on the KAK decomposition and template optimization occasionally yield good results but lead to worse results than the developed SMT approaches in most cases.

\section{Conclusion}\label{sec:conclusion}
\balance
In this work, we demonstrated the capability of semiconducting spins to support multiple two-qubit gates, yielding multiple universal quantum gate sets that can be used during quantum circuit adaptation to yield quantum circuits with a higher circuit fidelity or smaller qubit idle time.

The developed SMT approach is particularly well suited to deal with multiple two-qubit gates and yields a decrease in qubit idle time of up to 87\% and an increase in Hellinger fidelity of up to 40\% compared to direct basis translation.
Future research could include the development of suitable heuristics and the consideration of n-qubit gates.
\section*{Acknowledgment}
This work arose as a project from the Qiskit Advocate Mentorship Program in the Fall of 2021. The authors acknowledge the use of IBM Quantum Services for this work.
This work was partially funded by the Carl Zeiss foundation.
\renewcommand*{\bibfont}{\footnotesize}
\renewcommand*{\UrlFont}{\rmfamily}

\printbibliography
\end{document}